\documentclass[10pt,twocolumn,aps,pra,showpacs]{revtex4}

\usepackage{bbm}
		
\begin{document}

\title{Quantum search algorithm tailored to clause satisfaction problems}
\author{Avatar Tulsi\\
        {\small Department of Physics, IIT Bombay, Mumbai-400076, India}}

\email{tulsi9@gmail.com}

\begin{abstract}
Many important computer science problems can be reduced to clause
satisfaction problem. We are given $n$ Boolean variables $x_{k}$ and $m$
clauses $c_{j}$ where each clause is a function of values of some of
$x_{k}$'s. We want to find an assignment $i$ of $x_{k}$'s  for which all
$m$ clauses are satisfied. Let $f_{j}(i)$ be a binary function which is
$1$ if $j^{\rm th}$ clause is satisfied by the assignment $i$ else
$f_{j}(i) = 0$. Then the solution is $r$ for which $f(i=r) = 1$, where
$f(i)$ is the AND function of all $f_{j}(i)$'s. In quantum computing,
Grover's algorithm can be used to find $r$. A crucial component of this
algorithm is the selective phase inversion $I_{r}$ of the solution state
encoding $r$. $I_{r}$ is implemented by computing $f(i)$ for all $i$ in
superposition which requires computing AND of all $m$ binary functions
$f_{j}(i)$'s. Hence there must be coupling between the computation
circuits for each $f_{j}(i)$'s. In this paper, we present an alternative
quantum search algorithm which relaxes the requirement of such couplings.
Hence it offers implementation advantages for clause satisfaction
problems.
\end{abstract}


\maketitle

\section{INTRODUCTION}

Grover's algorithm is used to search an item satisfying certain properties
out of a database of $N$ items~\cite{grover}. Let the index $i$
($i\in\{0,1,\ldots,N-1\}$) denote these items. Consider a quantum system
of $n = \log_{2}N$ qubits with Hilbert space of dimension $N$, whose $N$
basis states is used to encode $N$ database items with one-to-one
correspondence. Grover's algorithm starts with the state $|\hat{0}\rangle$
in which all qubits are in $|0\rangle$ state. Then it applies
Walsh-Hadamard transformation $W$ which is nothing but application of
Hadamard gate on all qubits. After this, we get the state which is a
uniform superposition of all basis states, i.e.
\begin{equation}
W|\hat{0}\rangle = \frac{1}{\sqrt{N}}\sum_{i = 0}^{N-1}|i\rangle
\label{uniform}
\end{equation}
Grover's algorithm then successively iterates the Grover operator
\begin{equation}
\mathcal{G} = WI_{\hat{0}}WI_{r}
\end{equation}
on above state to get the basis state $|r\rangle$ encoding the solution of
search problem.

Here $I_{\hat{0}}$ and $I_{r}$ are the selective phase inversions of these
two states. Mathematically, they are written as
\begin{equation}
I_{\hat{0}} = \mathbbm{1} – 2|\hat{0}\rangle \langle \hat{0}|\ ;\
I_{r} = \mathbbm{1} – 2|r\rangle \langle r|\ . \label{selective}
\end{equation}
The number of iterations of $\mathcal{G}$ required by Grover's algorithm
is $(\pi/4)\sqrt{N}$ assuming that there is a unique solution $r$. This is
quadratically faster than classical search algorithms which take $O(N)$
time steps. Grover's algorithm is proved to be strictly
optimal~\cite{optimal}. Out of the two selective transformations,
$I_{\hat{0}}$ is easy to implement as we know the state $|\hat{0}\rangle$.
But we don't know the solution state $|r\rangle$ in advance and $I_{r}$ is
implemented using an \emph{oracle} transformation. Basically, for all
$|i\rangle$ in superposition, the oracle computes a binary function $f(i)$
which is $1$ if $|i\rangle = |r\rangle$ else $0$. Then the computed value
of $f(i)$ is used to selectively invert the phase of $|i = r\rangle$.

In the clause satisfaction problems, we have $n$ Boolean variables $x_{k}$
($k \in \{1,2,\ldots,n\}$) and each variable can take two values $0$ or
$1$. Let the index $i$ ($i \in \{0,1,\ldots,N-1\}$) denote the different
possible assignments of these variables where $N = 2^{n}$. We have $m$
clauses $c_{j}$ ($j \in\{1,2,\ldots,m\}$) where satisfaction of each
clause depends upon values of a subset of $n$ variables. Typically this
subset involves few number of variables. For example, in the widely
studied NP-complete problem of \textbf{3SAT}, a clause is satisfied if OR
of $3$ terms is $1$, where each term is either a Boolean variable $x_{k}$
or its negation $\bar{x_{k}}$ ($\bar{x_{k}} = 1$ if $x_{k} = 0$ else
$\bar{x_{k}} = 0$). The solution is a particular assignment $i =r$ which
satisfies all $m$ clauses.

Obviously, in such problems, the function $f(i)$ is AND of $m$ different
binary functions $f_{j}(i)$ where $f_{j}(i) = 1$ if $j^{\rm th}$ clause
$c_{j}$ is satisfied else $0$. To use Grover's algorithm, we need to
compute $f(i)$ for all $|i\rangle$ in superposition to implement $I_{r}$.
As $f(i)$ is AND of all $f_{j}(i)$'s there has to be some coupling between
individual computation circuits corresponding to each $f_{j}(i)$. These
couplings may add significantly to physical implementation challenges of
Grover's algorithm depending upon the kind of hardware that will be used
for quantum computing in future.

In this paper, we present an alternative quantum search algorithm which is
naturally tailored to such kind of problems. It relaxes the requirement of
coupling computation circuits for each $f_{j}(i)$'s as we don't need to
compute AND of all $f_{j}(i)$'s. In next section, we present the algorithm and we present its analysis in Section III. The analysis mainly uses the results of general quantum search algorithm presented in ~\cite{general}. We then discuss and conclude in Section IV.

\section{ALGORITHM}

Let $\mathcal{H}_{N}$ denote the Hilbert space of dimension $N = 2^{n}$ of
$n$ qubits where each qubit represents one of the given Boolean variables
$x_{k}$. We attach an ancilla qubit to this system and let
$\mathcal{H}_{2}$ denote the corresponding two-dimensional Hilbert space.
We work in the $2N$ dimensional joint Hilbert space $\mathcal{H} =
\mathcal{H}_{2} \otimes \mathcal{H}_{N}$.

For each clause $j$, we define an operator $\mathcal{D}_{j}$ which
computes the binary function $f_{j}(i)$ corresponding to $j^{\rm th}$
clause and then applies a controlled transformation on the ancilla qubit.
If $f_{j}(i) = 1$ then it leaves the ancilla qubit unchanged else it
applies the single-qubit operator $R_{m}$ on the ancilla qubit, where
\begin{equation}
R_{m} = exp(\imath \pi/m)|0\rangle \langle 0| + exp(-\imath
\pi/m)|1\rangle \langle 1|\ . \label{sqoperator}
\end{equation}
Obviously, the operator $\mathcal{D}_{j}$ acts only on $n'+1$ qubits where
$n'$ is the number of Boolean variables involved in $j^{\rm th}$ clause.
The extra one qubit is the ancilla qubit on which $\mathcal{D}_{j}$ does a
controlled operation. The eigenspectrum of $\mathcal{D}_{j}$ is of the
following form
\begin{eqnarray}
\mathcal{D}_{j}(|0\rangle|i\rangle) & = & exp[\imath \pi(1-
f_{j}(i))/m]|0\rangle|i\rangle \nonumber \\
\mathcal{D}_{j}(|1\rangle|i\rangle) & = & exp[-\imath \pi
(1-f_{j}(i))/m]|1\rangle|i\rangle \ .
\end{eqnarray}

Now consider the operator
\begin{equation}
\mathcal{D} = \mathcal{D}_{m}\mathcal{D}_{m-1}\cdots
\mathcal{D}_{2}\mathcal{D}_{1}
\end{equation}
which is basically a product of $\mathcal{D}_{j}$'s for all $j$ from $1$
to $m$. As each $\mathcal{D}_{j}$ commute with each other, being diagonal
in nature, the order of operators doesn't matter for implementing
$\mathcal{D}$. Here we note that each unitary operator $\mathcal{D}_{j}$
involves computation of $f_{j}(i)$ for only one clause and for
implementing $\mathcal{D}$, we don't need to couple any of them with each
other. We just successively apply $\mathcal{D}_{j}$ operators for all $j$
in any chosen order. It is easy to see that the eigenspectrum of
$\mathcal{D}$ is given by
\begin{eqnarray}
\mathcal{D}(|0\rangle|i\rangle) & = & exp[\imath \pi
u_{i}/m]|0\rangle|i\rangle \nonumber \\
\mathcal{D}(|1\rangle|i\rangle) & = & exp[-\imath \pi
u_{i}/m]|1\rangle|i\rangle \ .  \label{Dequation}
\end{eqnarray}
Here $u_{i} = \sum_{j=1}^{m}(1-f_{j}(i))$ is the total number of clauses
unsatisfied by the assignment $i$. As $u_{r} = 0$ (the solution satisfies
all clauses by definition), we see that $\mathcal{D}$ has a
two-dimensional degenerate eigenspace orthogonally spanned by
$|0\rangle|r\rangle$ and $|1\rangle|r\rangle$ with eigenvalue $1$.

Now we present the algorithm. \\ (1) Initially, we put all qubits
including the ancilla qubit in $|0\rangle$ state. \\  (2) Then we apply
Walsh-Hadamard transform $W_{n+1}$ on all $n+1$ qubits including the
ancilla qubit. We get the state $|+\rangle_{n+1} =
|+\rangle|+\rangle_{n}$, where $|+\rangle = 1/\sqrt{2}(|0\rangle
+|1\rangle)$ is the ancilla qubit state and $|+\rangle_{n} =
(1/\sqrt{N})\sum_{i}|i\rangle$ is the uniform superposition of all basis
states $|i\rangle$ of $\mathcal{H}_{N}$. \\ (3) We perform $q$ times
iteration of the operator $\mathcal{A} =
W_{n+1}I_{0,\hat{0}}W_{n+1}\mathcal{D}$ on the initial state
$|+\rangle_{n+1}$. Here $\mathcal{D}$ is given by Eq. (\ref{Dequation})
and $I_{0,\hat{0}}$ is the selective phase inversion of the
$|0\rangle|\hat{0}\rangle$ state. It is easy to see that
$W_{n+1}I_{0,\hat{0}}W_{n+1}$ is nothing but the selective phase inversion
$I_{+,n+1}$ of the initial state $|+\rangle_{n+1}$. So we can write the
operator $\mathcal{A}$ as
\begin{equation}
\mathcal{A} = I_{+,n+1}\mathcal{D} \label{Aquation}
\end{equation}
After $q$ iterations, we get the state $|w_{q}\rangle =
\mathcal{A}^{q}|+\rangle_{n+1}$. \\ (4)As we show in next section, there
exists a $q$ for which $|w_{q}\rangle$ is close to the state
$|+\rangle|r\rangle$ and measuring $|w_{q}\rangle$ will output the
solution $r$ with a significant probability. \\ (5) We repeat the entire
algorithm few times to get the solution with probability very close to
$1$.

\section{ANALYSIS}
To prove the working of algorithm, we need to prove that
$\mathcal{A}^{q}|+\rangle_{n+1}$ is close to $|+\rangle|r\rangle state$
for some $q$. To prove this, we prove the reverse operation, i.e. we prove
that $(\mathcal{A}^{\dagger})^{q}|+\rangle|r\rangle$ is close to the
initial state $|+\rangle_{n+1}$ state. From (\ref{Aquation}), we have
\begin{equation}
\mathcal{A}^{\dagger} = \mathcal{D}^{\dagger}I_{+,n+1} \ .
\end{equation}
Note that $I_{+,n+1}$ is a self-inverse transformation. The above operator
is just a special kind of the general quantum search operator $\mathcal{S}
= D_{s}I_{t}^{\phi}$ that we have analyzed in ~\cite{general}. We just
need to make the following substitutions in $\mathcal{S}$,
\begin{eqnarray}
|s\rangle \rightarrow |+\rangle|r\rangle, &\ &|t\rangle \rightarrow
|+\rangle_{n+1}, \nonumber \\
\mathcal{S} \rightarrow \mathcal{A}^{\dagger} &\ & \phi \rightarrow \pi\ .
\label{substitutions}
\end{eqnarray}
We refer the readers to original paper~\cite{general} for the details of
analysis as here we just use the results of analysis.

In the analysis of ~\cite{general}, by convention, the initial state
$|s\rangle$ is the eigenstate of $D_{s}$ with eigenvalue $1$. In case of a
degenerate eigenspace orthonormally spanned by the states $|s_{m}\rangle$,
the initial state is chosen according to Eq. (1) of ~\cite{general} as
\begin{equation}
|s\rangle = \frac{1}{\alpha}\sum_{m}\langle s_{m}|t\rangle|s_{m}\rangle\
,\ \ \alpha^{2} = \sum_{m}|\langle s_{m}|t\rangle|^{2}\ .
\label{initialstatedefine}
\end{equation}
Note that here the notation $m$ is used just to conform to the analysis of
~\cite{general}. It should not be confused with the notation of this paper
where $m$ denotes the number of clauses. In our case, $\mathcal{D}$ has a
$2$-dimensional degenerate eigenspace spanned by $|0\rangle|r\rangle$ and
$|1\rangle|r\rangle$ with eigenvalue $1$. As $|t\rangle = |+\rangle_{n+1}$
in our case, it is easy to check that in (\ref{substitutions}), the state
$|s\rangle$ becomes $|+\rangle|r\rangle$ just to satisfy
(\ref{initialstatedefine}). Also, it is easy to check that in our
algorithm,
\begin{equation}
\alpha = |\langle r|W|\hat{0}\rangle| = 1/\sqrt{N} \ .
\end{equation}

Using the analysis of ~\cite{general} further, we find that only two
eigenstates $|\lambda_{\pm}\rangle$ with the corresponding eigenvalues
$e^{\imath \lambda_{\pm}}$ of $\mathcal{A}^{\dagger}$ are relevant for our
algorithm as the state $|+\rangle|r\rangle$ is almost completely spanned
by them. These eigenvalues $\lambda_{\pm}$ are given by Eq. (12) of
~\cite{general}. In our case, this equation becomes
\begin{equation}
\lambda_{\pm} = \pm\frac{2}{B\sqrt{N}}(\tan \eta)^{\pm 1}\ \ \cot 2\eta =
\frac{\Lambda_{1}\sqrt{N}}{2B}\ .\label{solutions}
\end{equation}
where
\begin{equation}
B = \sqrt{1 + \Lambda_{2}}\ ,\ \Lambda_{p} = \sum_{\ell \neq
\psi,r}|\langle \ell|+\rangle_{n+1}|^{2}\cot^{p}\frac{\theta_{\ell}}{2}\ .
\label{definitionLambda}
\end{equation}

Here $|\ell\rangle$ denote the eigenstates of $\mathcal{D}$ with
corresponding eigenvalues $exp(\imath \theta_{\ell})$. As the sum is over
$|\ell\rangle \neq |\psi\rangle|r\rangle$ (here $|\psi\rangle$ is any
state of ancilla qubit) and $\theta_{\ell = \psi,r} = 0$ by convention, we
need to consider only those eigenstates with non-zero eigenphases
$\theta_{\ell}$ to evaluate this sum. Using Eq. (\ref{Dequation}), we find
that such eigenstates with corresponding eigenphases are
\begin{eqnarray}
|0\rangle|i \neq r\rangle &\ ,\ & exp[\imath \pi u_{i}/m] \nonumber \\
|1\rangle|i \neq r\rangle &\ ,\ & exp[-\imath \pi u_{i}/m]\ .
\end{eqnarray}
So we have $|\langle \ell|+\rangle_{n+1}|^{2} = 1/2N$ for all $\ell$ and
we get
\begin{equation}
\Lambda_{p} = \frac{1}{2N}\sum_{\ell \neq \psi,r}\cot^{p}\frac{\pi
u_{i}}{2m} \ .
\end{equation}
It is easy to check that $\Lambda_{1}$ vanishes because the contributions
from eigenstates $|0\rangle|i \neq r\rangle$ identically cancels that from
eigenstates $|1\rangle|i \neq r\rangle$. Also, if $N_{u}$ denotes the
total number of assignments $i$ which don't satisfy $u$ out of $m$
clauses, then we can write $\Lambda_{2}$ as
\begin{equation}
\Lambda_{2}=\frac{1}{N}\sum_{u=1}^{m}N_{u}\cot^{2}\frac{\pi u}{2m} \ .
\label{Blambda2}
\end{equation}

With $\Lambda_{1} = 0$, (\ref{solutions}) indicates that $\eta =
\frac{\pi}{4}$ and so
\begin{equation}
\lambda_{\pm} = \pm\frac{2}{B\sqrt{N}}\ .
\end{equation}
With $\eta = \pi/4$ and $\phi = \pi$, Eq. (23) and (24) of ~\cite{general}
gives us the initial state $|s\rangle$ and the effect of iterating
$\mathcal{S}$ on $|s\rangle$ in terms of two relevant eigenstates
$|\lambda_{\pm}\rangle$. With our substitutions (\ref{ substitutions}), we
get
\begin{equation}
|+,r\rangle = -\imath/\sqrt{2}[e^{\imath \lambda_{+}/2}|\lambda_{+}\rangle
- e^{\imath \lambda_{-}/2}|\lambda_{-}\rangle], \label{slambdapmexpansion}
\end{equation}
and
\begin{equation}
(\mathcal{A}^{\dagger})^{q}|+,r\rangle =  -\imath/\sqrt{2} [e^{\imath
q'\lambda_{+}}|\lambda_{+}\rangle - e^{\imath
q'\lambda_{-}}|\lambda_{-}\rangle],
 \label{stateexpand}
\end{equation}
where $q'  = q+\frac{1}{2}$.

For $q = q_{\rm m} \approx \pi/2|\lambda_{\pm}| = \pi B\sqrt{N}/4$, the
state $(\mathcal{A}^{\dagger})^{q_{\rm m}}|+,r\rangle$ is very close to
the state given by
\begin{equation}
1/\sqrt{2}(|\lambda_{+}\rangle + |\lambda_{-}\rangle).
\end{equation}
As shown in ~\cite{general}, the above state has an amplitude of $1/B$
with the target state $|t\rangle$, which in our case is the state
$|+\rangle_{n+1}$. So we have proved that
\begin{equation}
|\langle +|_{n+1}(\mathcal{A}^{\dagger})^{\pi B\sqrt{N}/4}|+,r\rangle| =
|\langle +,r|\mathcal{A}^{\pi B\sqrt{N}/4}|+\rangle_{n+1}| = 1/B.
\end{equation}
Thus one running of our algorithm succeeds in finding the solution with a
probability of $1/B^{2}$ and $O(B^{2})$ times running of our algorithm
will give the solution with a probability approaching $1$. So we need a
total of $O(\pi\sqrt{N}B^{3}/4)$ iterations of $\mathcal{A}$ to solve the
search problem.

In general, for clause satisfaction problems $B = O(1)$. For example, in
the case of 3SAT problem, we know that with probability very close to $1$,
a randomly picked assignment $i$ will not satisfy $m/8 \pm O(\sqrt{m})$
clauses. So the sum in (\ref{Blambda2}) is approximately
$\cot\frac{\pi}{16} \approx 4.96 = O(1)$.

We point out that our analysis of ~\cite{general} holds only if
$|\lambda_{\pm}| \ll \theta_{\rm min}$, where $\theta_{\rm min}$ is the
minimum eigenphase of $\mathcal{D}$ different from $0$. In our case,
$\theta_{\rm min} = \pi/m$ and as we found $|\lambda_{\pm}| =
O(1/\sqrt{N})$, this assumption is satisfied as long as $N \gg m^{2}$
which is true in almost all practical situations.

\section{DISCUSSION AND CONCLUSION}

We have presented a quantum search algorithm which naturally relates to the clause satisfaction problems. This algorithm allows us to find the solution using $O(\sqrt{N})$ oracle queries without any necessity of coupling the individual computation circuits corresponding to individual clause satisfactions. This algorithm has a potential to save the computational resources required to implement the oracle transformation $I_{r}$ in Grover`s search algorithm. 

In spirit, our algorithm is quite similar to that presented by Kato~\cite{kato}. There Kato has shown that implementation of $I_{\hat{0}}$, the selective phase inversion of the $|\hat{0}\rangle$ (all qubits in $|0\rangle$) state is not necessary for quantum search algorithms and this operator can be replaced by an operator made up of only single-qubit gates and so physically easier to implement. Here we have shown that implementation of the selective phase inversion of the solution state $I_{r}$, which is physically harder to implement, is not necessary for quantum search algorithms for clause satisfaction problems. Rather, it can be replaced by physically easier to implement operators.

We believe that similar ideas can be used for other kind of search problems also to design quantum search algorithms which are physically easier to implement.

\end{document}